\documentclass[aps,prb,showpacs,superscriptaddress,twocolumn]{revtex4-1}

\usepackage{amsmath,graphics}
\usepackage[next]{inputenc}
\usepackage[dvips]{epsfig}
\usepackage[colorlinks=true,citecolor=blue,linkcolor=blue]{hyperref}
\usepackage{bbm}
\usepackage{bbold}
\usepackage{booktabs}
\usepackage{multirow}
\usepackage{hhline}
\usepackage{graphicx}
\usepackage{amsmath}
\usepackage{multirow}
\usepackage[usenames, dvipsnames]{color}
\usepackage{color,soul}

\def\be{\begin{equation}}
\def\ee{\end{equation}}

\def\bi{\begin{itemize}}
\def\ei{\end{itemize}}
\def\bn{\begin{enumerate}}
\def\en{\end{enumerate}}
\def\bea{\begin{eqnarray}}
\def\eea{\end{eqnarray}}

\def\ba{\begin{array}}
\def\ea{\end{array}}
\def\bd{\begin{displaymath}}
\def\ed{\end{displaymath}}

\begin{document}

\title{Superconducting Phase-Diagram of H$_3$S under High Magnetic Fields}

\author{Shirin Mozaffari}
\affiliation{National High Magnetic Field Laboratory, Florida State University, Tallahassee, FL 32310, USA}

\author{Dan Sun}
\affiliation{Los Alamos National Laboratory, Los Alamos, NM 87545, USA}

\author{Vasily S. Minkov}
\affiliation{Max-Planck-Institut fuer Chemie, Hahn-Meitner Weg 1, 55128 Mainz, Germany}

\author{Dmitry Knyazev}
\affiliation{Max-Planck-Institut fuer Chemie, Hahn-Meitner Weg 1, 55128 Mainz, Germany}

\author{Jonathan B. Betts}
\affiliation{Los Alamos National Laboratory, Los Alamos, NM 87545, USA}

\author{Mari Einaga}
\affiliation{KYOKUGEN, Graduate School of Engineering Science, Osaka University, Machikaneyamacho 1-3, Toyonaka, Osaka 560-8531, Japan}

\author{Katsuya Shimizu}
\affiliation{KYOKUGEN, Graduate School of Engineering Science, Osaka University, Machikaneyamacho 1-3, Toyonaka, Osaka 560-8531, Japan}

\author{Mikhail I. Eremets}
\affiliation{Max-Planck-Institut fuer Chemie, Hahn-Meitner Weg 1, 55128 Mainz, Germany}

\author{Luis Balicas}
\affiliation{National High Magnetic Field Laboratory, Florida State University, Tallahassee, FL 32310, USA}

\author{Fedor F. Balakirev}\email{fedor@lanl.gov}
\affiliation{Los Alamos National Laboratory, Los Alamos, NM 87545, USA}

\begin{abstract}
	  We report the temperature dependence of the upper critical fields $\mu_0H_{c2}(T)$ of the high temperature superconductor H$_3$S under applied pressures of 155 and 160 GPa through the electrical resistance transition observed under DC and pulsed magnetic fields up to 65 T, a record high combination of fields and pressures. We find that $H_{c2}(T)$ generally follows the Werthamer, Helfand and Hohenberg (WHH) formalism at low fields, albeit with noticeable deviations upon approaching our experimental limit of $\mu_0H = 65$ T.  In fact, $H_{c2}(T)$ displays a remarkably linear dependence on temperature over an extended temperature range also found in multigap or in strongly-coupled superconductors.  The best fit of $H_{c2}(T)$ to the WHH formula yields a negligible value for the Maki parameter $\alpha$ and for spin-orbit scattering constant $\lambda_{\text{SO}}$. However, its behavior is relatively well-described by a model based on strong coupling superconductivity with a coupling constant $\lambda \sim 2$. Therefore, we conclude that H$_3$S behaves as a strong-coupled orbital-limited superconductor over the entire range of temperatures and fields used for our measurements.
\end{abstract}
%\begin{abstract}
	%We report the temperature dependence of the upper critical fields $\mu_0H_{c2}(T)$ of the high temperature superconductor H$_3$S under applied pressures of 155 and 160 GPa through the electrical resistance transition observed under DC and pulsed magnetic fields up to 65 T, a record high combination of field and pressure. The best fit of $H_{c2}(T)$ to the WHH formula (weak coupling) yields a negligible value for the Maki parameter $\alpha$ and for spin-orbit scattering constant $\lambda_{\text{SO}}$. Instead, we find that $H_{c2}(T)$ generally follows the strong electron-phonon coupling  formalism with with electron-phonon coupling constant $\lambda\sim3$.  In fact, $H_{c2}(T)$ displays a remarkably linear dependence on temperature over an extended temperature range also found in some multigap and strongly-coupled superconductors. Therefore, we conclude that H$_3$S behaves as an orbital-limited superconductor over the entire range of temperatures and fields used for our measurements.
%\end{abstract}
\date{\today}

\maketitle
%\section{Introduction \label{intro}}
The ongoing scientific quest to stabilize superconductivity at room temperature led to the discovery of superconductivity, with a very high critical temperature $T_c$ = 203 K, in sulfur hydride H$_3$S under high pressures of $p$ = 155 GPa \cite{MEHS}. H$_3$S along with other hydrides\cite{LaH,Somayazulu_2019} seems to be the closest compound, so far, to metallic hydrogen, which is predicted to be a high temperature superconductor\cite{Ashcroft,Ginzburg,Maksimov}.
H$_3$S forms as the result of the chemical instability of H$_2$S under high pressures, where H$_2$S decomposes into elemental sulfur S and H$_3$S\cite{Ma1,Duan,Errea_2015}. The structure of H$_3$S is believed to be body-centered cubic \textit{Im-$\bar{3}$m} over the pressure range of 92 - 173 GPa \cite{Einaga}, and characterized by H atoms situated midway between two  body-centered S atoms. The pressure as a function of temperature phase diagram of H$_3$S is asymmetric, meaning that $T_c$ shows a sharp increase from 95 K to 203 K in the pressure range of 110 - 155 GPa, but decreases with further increasing the pressure\cite{MEHS}.

The superconductivity in H$_3$S is believed to be conventional, in which pairing is mediated by phonons, with high-frequency optical modes due to the motion of hydrogen\cite{MEHS,Kresin,phonon1,phonon2,Kresin2}. The high $T_c$ arises from both the metallization of the strongly covalent bonds in H$_3$S\cite{Igor} and the high phonon frequencies displayed by its light elements\cite{Pickett}. Band structure calculations indicate that H$_3$S is a multiband metal \cite{Bianconi} having a large Fermi surface (broad energy dispersive bands) as the result of the hybridization between the H 1\textit{s} and the 3\textit{p} orbitals of sulfur\cite{Duan,Errea_2015,Pickett,Heil,Flores,phonon2}.
Band structure calculations also yield small Fermi surface pockets for the high-$T_c$ phase of H$_3$S \cite{Bianconi}.

Despite the very high $T_c$s reported for H$_3$S, the studies of $H_{c2}(T)$ are limited to a narrow range of temperatures close to $T_c$ due to the inherent experimental difficulties in performing ultra-high pressure measurements under very high magnetic fields. The behavior of $H_{c2}(T)$ provides valuable information such as an estimation of the Cooper pair coherence length, the strength of the electron-phonon coupling, the role of the spin-orbit coupling, and the dominant mechanism breaking the Copper pairs.

In type-II superconductors, the orbital and the spin-paramagnetic pair-breaking effects are the two main mechanisms depairing electrons under high magnetic fields. The orbital pair breaking effect explains the suppression of superconductivity via the formation of Abrikosov vortices\cite{Abrikosov} in the presence of a field. The superconductivity is suppressed when the kinetic energy associated with vortex currents exceeds the condensation energy of the paired electrons. When the magnetic field approaches a critical value $H_{c2}^{\text{orb}}=\phi_0/2 \pi \xi^2$, referred to as the orbital limiting field, the vortex cores overlap and the system returns to the normal state. Here, $\phi_0$ is the flux quantum and $\xi$ is the coherence length. The spin-paramagnetic effect explains the effect of the magnetic field on the spin of the electron based on the classical work by Clogston\cite{Clogston} and Chandrasekhar\cite{Chandrasekhar}. When the magnetic energy exceeds the superconducting gap, i.e. $ 1/2  \chi_p H_p^2 = 1/2 N(E_F) \Delta^2$, superconductivity will be suppressed, where $\chi_p$, $N(E_F)$, and $\Delta$ are the normal state paramagnetic susceptibility, the density of states at the Fermi level, and the superconducting gap, respectively. If we only consider the spin-paramagnetic effect, the zero-temperature Pauli limiting field (Chandrasekhar-Clogston limit) for a weakly coupled superconductor is approximately $\mu_0 H_p(0) = 1.86~T_c$, where $2\Delta(T=0)\sim 3.52~k_BT_c$ (for a conventional BCS superconductor) and $\chi_p=g\mu_B^2 N(E_F)$ ($g$ is the Land\'{e} $g$ factor and $\mu_B$ is the Bohr magneton). $H_p$ can be renormalized by the strength of the electron phonon-coupling or by electronic correlations. In many superconductors $H_{c2}(T)$  is affected by both pair-breaking mechanisms. The Maki\cite{Maki} parameter $\alpha = \sqrt{2}H_{c2}^{\text{orb}}(0)/H_p(0)$ is a measure of the relative strength between the orbital and the paramagnetic pair breaking mechanisms for a given type-II superconductor. The WHH theory\cite{WHH} includes both pair-breaking effects through the Maki parameter and the spin-orbit constant.

Here, we report measurements of the upper critical field in H$_3$S samples under extremely high pressures in high magnetic fields. The ultrahigh pressure can only be obtained by a diamond anvil cell (DAC), however, measurements on samples contained by a DAC are quite challenging, particularly in pulsed fields due to the narrow magnet bore and large induced currents which heat the cell. The small diameter of the DAC\cite{MEHS} employed in our study allows us to perform transport measurements up to 65 T without heating the sample significantly. Our pulsed field measurements under $p>150$ GPa is a significant pressure increase over latest pulsed field achievement of 4 GPa\cite{Jacques}.
We find that H$_3$S is an orbital limited superconductor over the entire temperature range and likely a multigap or strongly-coupled superconductor.

%\section{Experiments}\label{Experiments}
Two samples of H$_3$S were synthesized \textit{in-situ} inside DACs using two different techniques. The first sample was prepared from condensed liquid H$_2$S via disproportionation reaction as described in Ref.~[\onlinecite{MEHS}], the final pressure inside the sample was as high as 160 GPa. We refer to this sample as the 160 GPa sample. The second sample was synthesized directly from elemental sulfur and hydrogen at high pressure. For this sample a small piece of elemental sulfur (purity of 99.98\%) with a lateral dimension of about 20 $\mu$m and a thickness of $\sim 3-5$ $\mu$m was placed in a DAC having diamonds with culets of $60-70$ $\mu$m.
Excess hydrogen was introduced in the DAC at a gas pressure of  $ \sim 130 - 150$ MPa. Subsequently, the DAC was pressurized up to 150 GPa and then heated up to 1000 K with a YAG laser at room temperature to initiate the chemical reaction. Pressure increased slightly to 155 GPa after the synthesis. The vibrational properties of the pressurized initial reactants and of the synthesized products were probed using a triple grating Raman spectrometer equipped by a HeNe laser with a wavelength of 633 nm.
Sputtered gold electrodes were thoroughly isolated from the metal gasket by a layer made from magnesium oxide, calcium fluorite and an epoxy glue mixture. This layer also prevented hydrogen penetration into the rhenium gasket.  The pressure was estimated using the Raman shift of stressed diamond\cite{ME-Raman} and the H$_2$ vibron's wavenumber, previously calibrated in a separate experiment\cite{ME-dense}. Both scales indicated a pressure of 155 GPa after the laser heating.  Under continuous fields up to $\mu_0H=35$ T the resistance of the sample was measured using a commercial AC resistance bridge. A custom made Lock-In amplifier was used under pulsed fields up $\mu_0H=65$ T.

\begin{figure}
	\includegraphics[width=\linewidth]{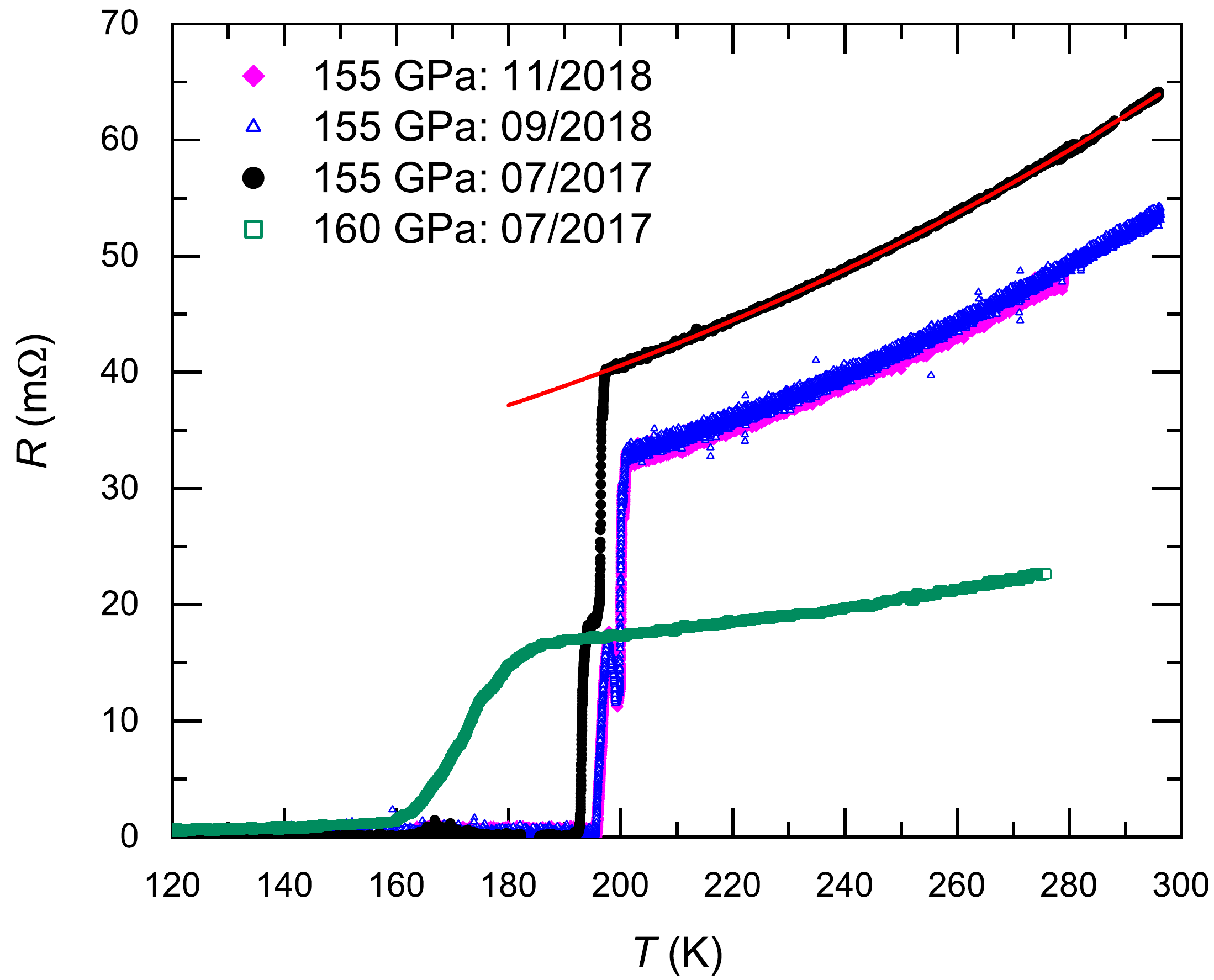}
	\caption{(Color online) Temperature dependence of resistance of the H$_3$S samples under 155 GPa and 160 GPa pressures. Red line is a fit to $R=R_0+AT^2$ indicating Fermi liquid behavior. Note that $T_c$ for the 155 GPa sample shifts slightly to larger values over time.} \label{RvsT}
\end{figure}

%\section{Results and Discussion}
\begin{figure}
	\centering
	\includegraphics[width= 7 cm]{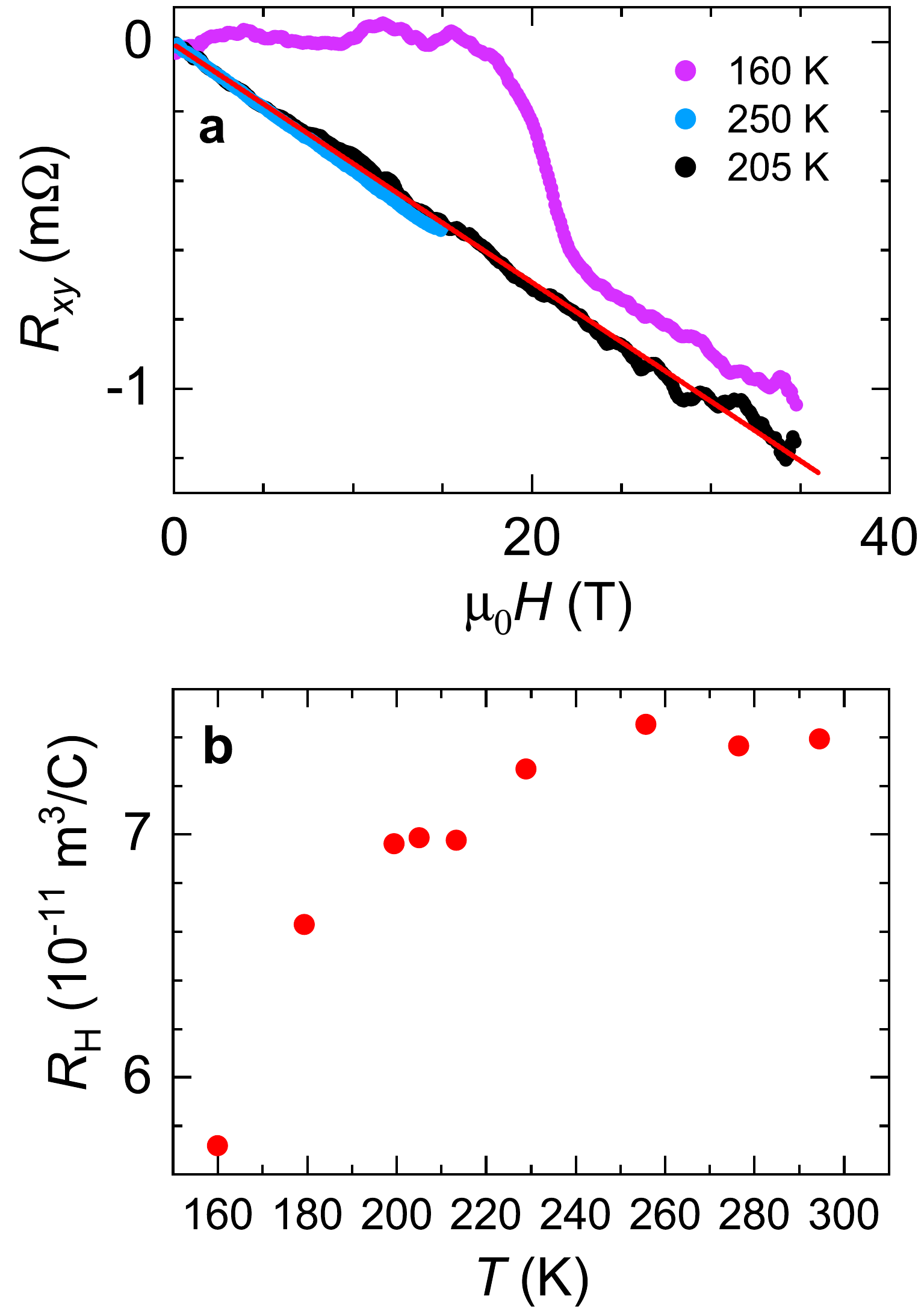}
		\caption{(Color online) (a) Hall resistance $R_{xy}$ as a function of the magnetic field $\mu_0H$ at a temperatures $T = 205$ snd 250 K. Red line is a linear fit. (b) Hall coefficient $R_{H}$ as a function of $T$ for the H$_3$S sample pressurized up to 155 GPa. } \label{Hall}
\end{figure}
Figure ~\ref{RvsT} displays the resistance as a function of the temperature $T$ for H$_3$S samples pressurized up to 155 GPa and 160 GPa. The onset, i.e. first deviation from normal state resistivity, of the superconducting transitions are estimated to be $T_c=201$ K for the sample under 155 GPa, and $T_c=174$ K for the other sample under 160 GPa.
The drop in $T_c$ for the higher pressure sample is consistent with the results in Ref.~[\onlinecite{MEHS}]. The width of these transitions are $\Delta T_c=5.5$ K and 26 K for the samples under 155 GPa and 160 GPa, respectively indicating that the latter sample is less homogeneous, or that it is subjected to stronger pressure gradients. Both samples display Fermi-liquid behavior in the normal state as indicated by the red line which corresponds to a fit to $R(T)=R_0+AT^2$.
The value of $T_c$ for the sample under 155 GPa increases slightly over time, indicating an evolution towards a more homogeneous sample or weaker pressure gradients. Notice that this sample displays a step around 195 K in the superconducting transition indicating that its superconducting state indeed is inhomogeneous. For samples synthesized within the confines of the DAC a certain degree of inhomogeneity is inevitable.

Figure~\ref{Hall}~(a) displays the Hall resistance $R_{xy}$ for the sample under $p=155$ GPa  as a function of magnetic field $\mu_0H$ at 205 K or near the superconducting transition. The $R_{xy}(\mu_0H)$ does not reveal any non-linearity as a function of $\mu_0H$ all the way up to 35 T. This indicates that near and above $T_c$ the electrical transport in H$_3$S is dominated by single electron-like carrier pocket despite the fact that H$_3$S was predicted to be a multiband metal\cite{Bianconi}. Most likey, this indicates that electrons display a considerably higher carrier mobility than holes at these temperatures. The temperature dependence of the Hall coefficient $R_H$ is shown in Fig.~\ref{Hall}~(b). We observed a noticeable decrease in $R_{H}(T)$ as the temperature is decreased, corresponding to an increase in the effective Hall density $n_H=1/eR_{H}$. The decrease in the Hall coefficient at low $T$s could come from a decrease in electron mobility, a progressive increase in hole mobility, or from an evolution in their relative densities. If one assumed that only one band contributed to the transport of carriers, the Hall coefficient would yield an electron density $n=1/eR_H=8.5\times10^{22}$ cm$^{-3}$ at room temperature, which is relatively close to known values for transition metals such as copper. At 205 K, the carrier concentration would increase to $n=8.9\times10^{22}$ cm$^{-3}$.\par
\begin{figure}
	\includegraphics[width=8.6 cm ]{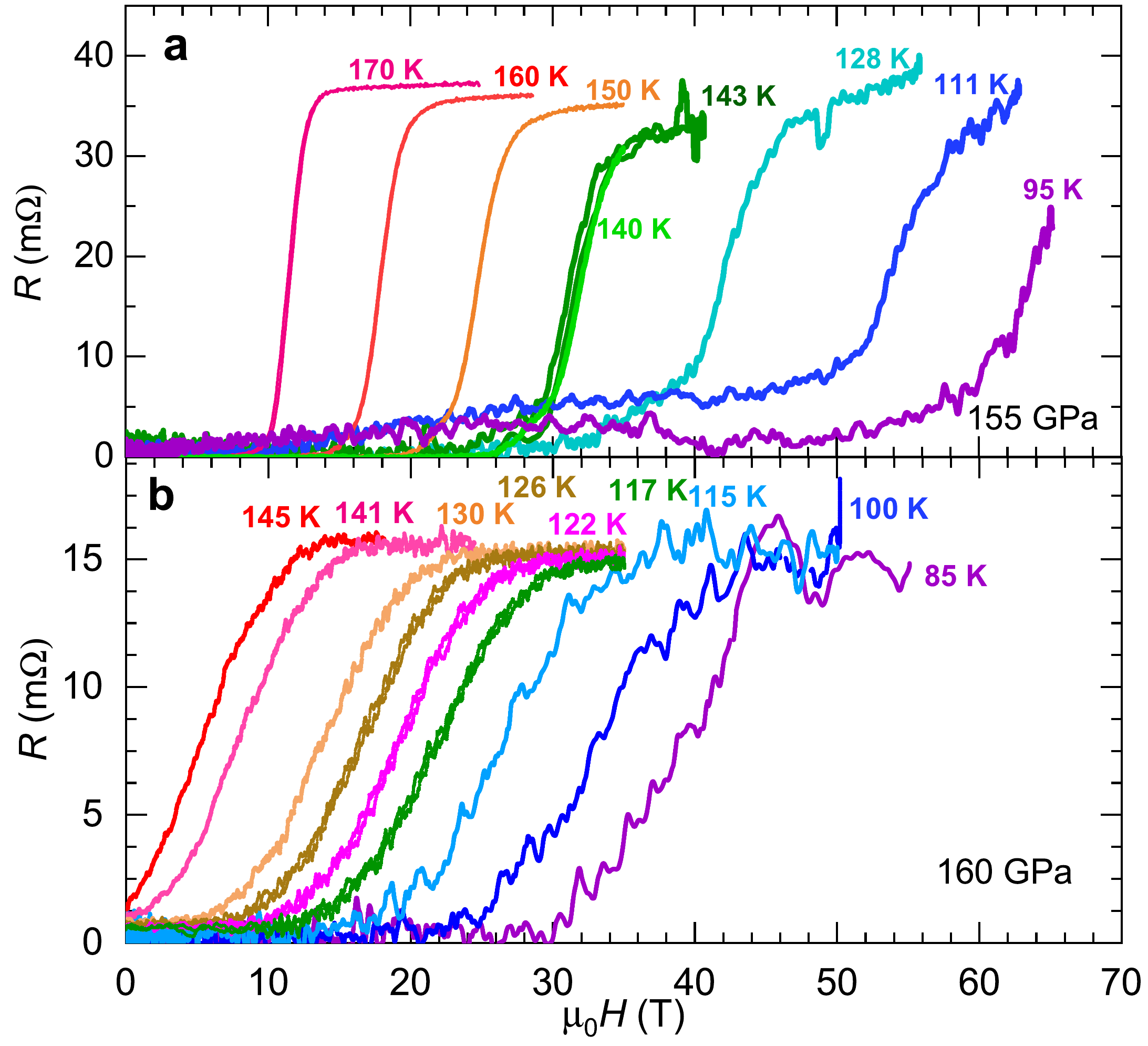}
	\caption{(Color online) Resistance as a function of field $\mu_0H$ for the H$_3$S samples under (a) $p = 155$ GPa and (b) 160 GPa  and for several temperatures.
		This data was collected under continuous and pulsed field.} \label{RvsH}
	%Black dashed lines indicate the definition of $H_{c2}$ and $H^\ast$.
\end{figure}
To determine the temperature dependence of $H_{c2}$, we measured the isothermal resistance as a function of the magnetic field $\mu_0H$ at selected temperatures ranging from $T = 60$ K to 200 K. Figures~\ref{RvsH}~(a) and (b) show the magnetic field dependence of the resistance for the 155 GPa and 160 GPa samples under fixed temperatures. At each temperature the resistance changes from zero to a finite value as $\mu_0H$ increases due to the field-induced suppression of superconductivity.  $H_{c2}$ is defined as the intersection between an extrapolation of the resistance of the normal state and a line having the slope the resistive transition at its middle point \cite{Budko}. The same criterium is used across DC and pulsed field traces.

The resistive onset has been found to match thermodynamic $H_{c2}$ obtained by other experimental probes\cite{Mielke_2001,Altarawneh_2008,Grissonnanche2014}. This criterion prevents the contribution of possible superconducting vortex related phases. The resistive transition shifts to higher fields as the temperature is lowered while broadening slightly. The $H_{c2}(T)$ values obtained from pulsed field measurements agree with the curvature and the values extracted under DC fields. At very high pulsed magnetic fields the self-heating of the metallic DACs by the induced currents becomes noticeable at temperatures below 60 K.

\par
\begin{figure}
	\includegraphics[width=8.6 cm]{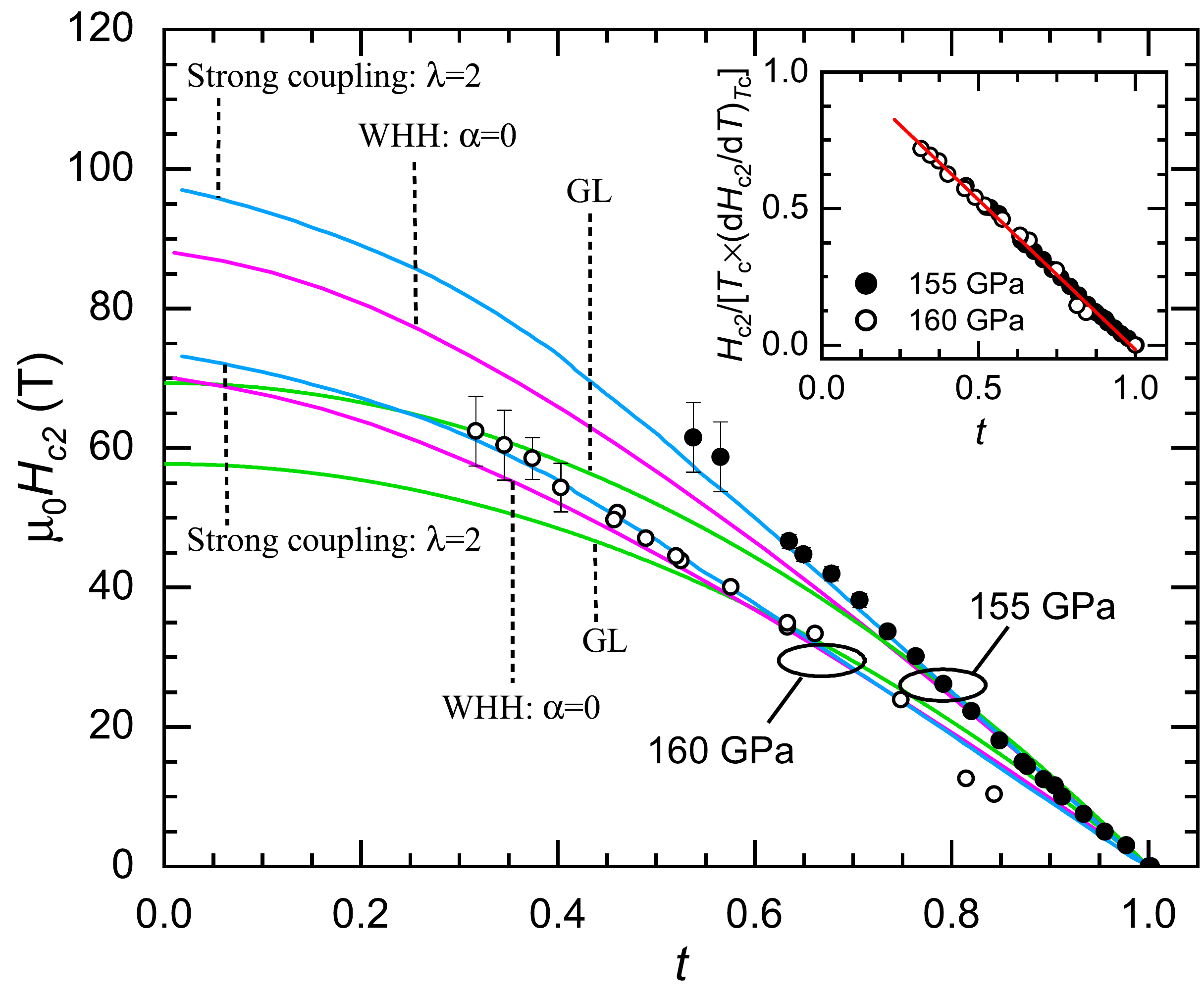}
	\caption{(Color online) Upper critical fields $H_{c2}$ as a function of reduced temperature $t=T/T_c$ for the H$_3$S samples under 155 GPa and 160 GPa pressure. Solid lines are fits to theoretical models: Ginzburg-Landau (GL), Werthamer, Helfand and Hohenberg (WHH), and strong strong-coupling model using the coupling strength parameter $\lambda=2$ (reproduced from a prediction based on strong-coupling theory \cite{Boulaevskii_1988,Thomas1996}). Inset: H$_3$S upper-critical fields for two different pressures plotted in reduced variables. Red line is a linear fit.} \label{hvst}
\end{figure}
Figures~\ref{hvst} shows the extracted $H_{c2}$ as a function of $T$ for both samples under $p= 155$ GPa and 160 GPa as a function of the reduced temperature $t=T/T_c$. To avoid the influence of possible superconducting vortex phases, we plot just the onset of their resistive transition. These phase boundaries were obtained from both isothermal field scans and from temperature scans under fixed magnetic fields. For both samples $H_{c2}(T)$ increases almost linearly upon decreasing the temperature under fields all the way up to 60 T yielding the slopes $|dH_{c2}/dT|_{T_{c}} = 0.62$ and 0.45 T/K for the samples under 155 GPa and 160 GPa, respectively. The linear dependence of $H_{c2}$ over an extended range of temperatures was also observed in the two-band superconductor MgB$_2$\cite{MgB2,AlexG} and in the multigap Fe-pnictide superconductors for fields along certain, but not for \textit{all} crystallographic orientations, and claimed to result from the orbital limiting effect \cite{Luis-non-Ising,FFB,NLWang}. Multigap effects might explain our phase-diagram for H$_3$S in analogy with those systems. However, multiband scenario contrast with the linear in field Hall data in Fig.~\ref{Hall}~(a), which indicates conduction dominated by a single type of carrier. Measurements of, for example, the penetration depth as a function of $T$ should help clarify the nature of the superconducting state in this system.
The estimated weak coupling Pauli limiting fields for the samples under 155 GPa and 165 GPa are $H_p(0)\simeq$~370 T and 344 T, respectively. In the dirty limit, the  $H_{c2}(0)$ values derived from the slope of the $H$-$T$ phase boundary at $T_c$ are, $H_{c2}^{orb}(0)=0.69~|dH_{c2}/dT|_{T_c} \times T_c=$ 88 T for the sample under 155 GPa and 74 T for the one under 160 GPa.  These latter values are much closer to the extrapolation of our experimental $H_{c2}(T)$ towards zero-temperature, when compared to the Pauli limiting values which differ by a factor of $\sim 5$, thus indicating that H$_3$S is an orbital limited superconductor.

Subsequently, we fit our experimentally determined phase-boundary to two different expressions in order to estimate the value of $H_{c2}$ at $T=0$ K.
First, the conventional Ginzburg-Landau (GL) expression:
\begin{equation}
H_{c2}(T)=\dfrac{\phi_0}{2\pi \bar{\xi}(0)^2}\left(1-t^2 \right),
\end{equation}
where $\phi_0$ is the quantum of flux and $\bar{\xi}(0)$ is an average Ginzburg-Landau coherence length at $T=0$.

And second WHH formalism where the temperature dependence of $H_{c2}$ defined by orbital and spin-paramagnetic effects in the dirty limit is given by \cite{WHH}:
\begin{equation}\label{WHH-eq}
\begin{split}
\ln \left( \dfrac{1}{t} \right) &= \sum_{\nu=-\infty}^{\infty}  \Bigg\{  \dfrac{1}{|2 \nu +1|}- \Bigg[ |2 \nu +1|
\\
& + \dfrac{\bar{h}}{t} + \dfrac{(\alpha \bar {h}/t)^{2}}{|2 \nu +1|+(\bar {h}+ \lambda_{so})/t}\Bigg]^{-1} \Bigg\}
\end{split}
\end{equation}
where,  $\bar {h} = (4/\pi^2)[H_{c2}(T)/(-dH_{c2}/dT)_{T_c}]$, $\alpha$ is the Maki parameter, and $\lambda_{\text{SO}}$ is the spin-orbit constant.

At temperatures near $T_c$, our experimental $H_{c2}$ values seem to follow the usual GL-expression, but at low temperatures they deviate from it considerably. This is particularly apparent for the sample under 160 GPa, for which we were able to reach smaller $t$ values. Comparing the upper critical fields to the WHH expression yields very small values for $\alpha$ which points again to rather small orbital limiting fields relative to the Pauli limiting ones, with no apparent saturation in $H_{c2}(t)$ at low $t$s. We obtained the best fits to the WHH formula for $\alpha$ values ranging between 0.0 and 0.3 for the sample under 155 GPa and $\alpha=0-0.2$ for the sample under 160 GPa. For such small values of $\alpha$, the WHH formula is almost insensitive to the strength of the spin-orbit interaction. 
As for the effect of pressure on the H$_3$S samples, our experimental upper critical fields do not reveal any change in the relative strength of both pair breaking mechanisms between both pressures. As seen in the inset of Fig. \ref*{hvst}, the superconducting phase diagrams for both samples fall into a single curve when plotted in reduced units. What is remarkable is the linearity of the phase-boundary as a function of $t$ all the way down to $t \sim 0.25$. Again, this resembles behavior reported, for example, for the Fe based superconductors \cite{NLWang}.

This deviation from conventional WHH could be also explained by fact that it was largely formulated in the weak limit of electron-phonon coupling constant $\lambda \ll 1$. Recent \textit{ab initio} calculations report substantial enhancement of $\lambda$ in H$_3$S due to the proximity to a  structural instability between \textit{Im-$\bar{3}$m} and $R3m$ crystal symmetries\cite{Errea2016}. As phonon frequencies $\omega_p$ soften near this instability, Cooper pairing is weakened at elevated temperatures by thermal phonons, while it remains robust in $T \rightarrow 0$ limit, leading to a relative enhancement of $H_{c2}$ at low temperatures.   The calculated  $\lambda \sim 2$ \cite{Errea_2015,Errea2016} would correspond to a nearly linear $H_{c2}(T)$ for $t > 0.25$  within the framework of the strong coupling model\cite{Boulaevskii_1988,Thomas1996}, which was also found to be applicable in an unusually strong electron-phonon coupling case in Bi-III \cite{Brown_2018}. Coincidentally, the strong coupling model also predicts a very high $T_c \sim$300-400 K for a lattice containing light atoms like H\cite{Boulaevskii_1988}. Our experimental $H_{c2}(T)$ values display a much better agreement with the temperature dependence predicted by the strong coupling model\cite{Boulaevskii_1988} with an electron-phonon coupling constant $\lambda=2$. We obtained the superconducting coherence length by extrapolating the strong-coupling fit to zero-temperature. This extrapolation of $H_{c2}(T)$s yields $H_{c2}=97$ T and 73 T for the H$_3$S samples pressurized up to 155 GPa and 160 GPa, respectively. Thus, we find a coherence length $\xi=1.84$ nm and of 2.12 nm for the samples under 155 GPa and 160 GPa, respectively.

%\section{Conclusion}\label{Conclusion}
In summary, we have investigated the temperature dependence of the upper critical fields of H$_3$S under magnetic fields up to 65 T.
At lower fields the phase-boundary separating normal and superconducting states is relatively well described by the Werthamer, Helfand and Hohenberg formula. Pronounced
deviations from the WHH formula are observed at lower temperatures due to the linearity of $H_{c2}(T)$. Overall, the phase boundary between superconducting and metallic states
indicates that the orbital-effect suppresses superconductivity over the entire temperature range. The linearity of $H_{c2}(T)$ observed over an extended range of reduced temperatures suggests that this system might indeed be a multigap superconductor as predicted theoretically and in analogy with similar results for the Fe based superconductors.  Alternatively, enhanced electron-phonon coupling and softening of hydrogen vibrational modes in the vicinity of structural instability could also explain the deviation from WHH predictions. We extract values for the coherence length $\xi$ ranging between  $\xi= 18.4~\text\AA$ and 21.2 \AA. Above the transition temperature the Hall-effect remains linear up to high magnetic fields indicating that H$_3$S is a very good metal and that at high temperatures carrier conduction is dominated by electrons.

%\section*{Acknowledgment}
 We acknowledge A. V. Gurevich and H. H. Wen for useful discussions. This work was performed at the National High Magnetic Field laboratory, which is supported by National Science Foundation Cooperative Agreement No. DMR-1644779, and the State of Florida. S.M. acknowledges support from the FSU Provost Postdoctoral Fellowship Program. L.B. is supported by DOE-BES through award DE-SC0002613. M.I.E. is thankful to  the Max Planck community for the invaluable support and U. P\"{o}schl for the constant encouragement.

\hfill

\section*{References}

\bibliography{Refs}

\begin{thebibliography}{40}%
\makeatletter
\providecommand \@ifxundefined [1]{%
 \@ifx{#1\undefined}
}%
\providecommand \@ifnum [1]{%
 \ifnum #1\expandafter \@firstoftwo
 \else \expandafter \@secondoftwo
 \fi
}%
\providecommand \@ifx [1]{%
 \ifx #1\expandafter \@firstoftwo
 \else \expandafter \@secondoftwo
 \fi
}%
\providecommand \natexlab [1]{#1}%
\providecommand \enquote  [1]{``#1''}%
\providecommand \bibnamefont  [1]{#1}%
\providecommand \bibfnamefont [1]{#1}%
\providecommand \citenamefont [1]{#1}%
\providecommand \href@noop [0]{\@secondoftwo}%
\providecommand \href [0]{\begingroup \@sanitize@url \@href}%
\providecommand \@href[1]{\@@startlink{#1}\@@href}%
\providecommand \@@href[1]{\endgroup#1\@@endlink}%
\providecommand \@sanitize@url [0]{\catcode `\\12\catcode `\$12\catcode
  `\&12\catcode `\#12\catcode `\^12\catcode `\_12\catcode `\%12\relax}%
\providecommand \@@startlink[1]{}%
\providecommand \@@endlink[0]{}%
\providecommand \url  [0]{\begingroup\@sanitize@url \@url }%
\providecommand \@url [1]{\endgroup\@href {#1}{\urlprefix }}%
\providecommand \urlprefix  [0]{URL }%
\providecommand \Eprint [0]{\href }%
\providecommand \doibase [0]{http://dx.doi.org/}%
\providecommand \selectlanguage [0]{\@gobble}%
\providecommand \bibinfo  [0]{\@secondoftwo}%
\providecommand \bibfield  [0]{\@secondoftwo}%
\providecommand \translation [1]{[#1]}%
\providecommand \BibitemOpen [0]{}%
\providecommand \bibitemStop [0]{}%
\providecommand \bibitemNoStop [0]{.\EOS\space}%
\providecommand \EOS [0]{\spacefactor3000\relax}%
\providecommand \BibitemShut  [1]{\csname bibitem#1\endcsname}%
\let\auto@bib@innerbib\@empty
%</preamble>
\bibitem [{\citenamefont {Drozdov}\ \emph {et~al.}(2015)\citenamefont
  {Drozdov}, \citenamefont {Eremets}, \citenamefont {Troyan}, \citenamefont
  {Ksenofontov},\ and\ \citenamefont {Shylin}}]{MEHS}%
  \BibitemOpen
  \bibfield  {author} {\bibinfo {author} {\bibfnamefont {A.~P.}\ \bibnamefont
  {Drozdov}}, \bibinfo {author} {\bibfnamefont {M.~I.}\ \bibnamefont
  {Eremets}}, \bibinfo {author} {\bibfnamefont {I.~A.}\ \bibnamefont {Troyan}},
  \bibinfo {author} {\bibfnamefont {V.}~\bibnamefont {Ksenofontov}}, \ and\
  \bibinfo {author} {\bibfnamefont {S.~I.}\ \bibnamefont {Shylin}},\ }\href
  {https://doi.org/10.1038/nature14964} {\bibfield  {journal} {\bibinfo
  {journal} {Nature}\ }\textbf {\bibinfo {volume} {525}},\ \bibinfo {pages}
  {73} (\bibinfo {year} {2015})}\BibitemShut {NoStop}%
\bibitem [{\citenamefont {Drozdov}\ \emph {et~al.}(2018)\citenamefont
  {Drozdov}, \citenamefont {Kong}, \citenamefont {Minkov}, \citenamefont
  {Besedin}, \citenamefont {Kuzovnikov}, \citenamefont {Mozaffari},
  \citenamefont {Balicas}, \citenamefont {Balakirev}, \citenamefont {Graf},
  \citenamefont {Prakapenka}, \citenamefont {Greenberg}, \citenamefont
  {Knyazev}, \citenamefont {Tkacz},\ and\ \citenamefont {Eremets}}]{LaH}%
  \BibitemOpen
  \bibfield  {author} {\bibinfo {author} {\bibfnamefont {A.~P.}\ \bibnamefont
  {Drozdov}}, \bibinfo {author} {\bibfnamefont {P.~P.}\ \bibnamefont {Kong}},
  \bibinfo {author} {\bibfnamefont {V.~S.}\ \bibnamefont {Minkov}}, \bibinfo
  {author} {\bibfnamefont {S.~P.}\ \bibnamefont {Besedin}}, \bibinfo {author}
  {\bibfnamefont {M.~A.}\ \bibnamefont {Kuzovnikov}}, \bibinfo {author}
  {\bibfnamefont {S.}~\bibnamefont {Mozaffari}}, \bibinfo {author}
  {\bibfnamefont {L.}~\bibnamefont {Balicas}}, \bibinfo {author} {\bibfnamefont
  {F.}~\bibnamefont {Balakirev}}, \bibinfo {author} {\bibfnamefont
  {D.}~\bibnamefont {Graf}}, \bibinfo {author} {\bibfnamefont {V.~B.}\
  \bibnamefont {Prakapenka}}, \bibinfo {author} {\bibfnamefont
  {E.}~\bibnamefont {Greenberg}}, \bibinfo {author} {\bibfnamefont {D.~A.}\
  \bibnamefont {Knyazev}}, \bibinfo {author} {\bibfnamefont {M.}~\bibnamefont
  {Tkacz}}, \ and\ \bibinfo {author} {\bibfnamefont {M.~I.}\ \bibnamefont
  {Eremets}},\ }\href {https://arxiv.org/abs/1812.01561} {\bibfield  {journal}
  {\bibinfo  {journal} {arXiv:1812.01561}\ } (\bibinfo {year}
  {2018})}\BibitemShut {NoStop}%
\bibitem [{\citenamefont {Somayazulu}\ \emph {et~al.}(2019)\citenamefont
  {Somayazulu}, \citenamefont {Ahart}, \citenamefont {Mishra}, \citenamefont
  {Geballe}, \citenamefont {Baldini}, \citenamefont {Meng}, \citenamefont
  {Struzhkin},\ and\ \citenamefont {Hemley}}]{Somayazulu_2019}%
  \BibitemOpen
  \bibfield  {author} {\bibinfo {author} {\bibfnamefont {M.}~\bibnamefont
  {Somayazulu}}, \bibinfo {author} {\bibfnamefont {M.}~\bibnamefont {Ahart}},
  \bibinfo {author} {\bibfnamefont {A.~K.}\ \bibnamefont {Mishra}}, \bibinfo
  {author} {\bibfnamefont {Z.~M.}\ \bibnamefont {Geballe}}, \bibinfo {author}
  {\bibfnamefont {M.}~\bibnamefont {Baldini}}, \bibinfo {author} {\bibfnamefont
  {Y.}~\bibnamefont {Meng}}, \bibinfo {author} {\bibfnamefont {V.~V.}\
  \bibnamefont {Struzhkin}}, \ and\ \bibinfo {author} {\bibfnamefont {R.~J.}\
  \bibnamefont {Hemley}},\ }\href {\doibase 10.1103/PhysRevLett.122.027001}
  {\bibfield  {journal} {\bibinfo  {journal} {Phys. Rev. Lett.}\ }\textbf
  {\bibinfo {volume} {122}},\ \bibinfo {pages} {027001} (\bibinfo {year}
  {2019})}\BibitemShut {NoStop}%
\bibitem [{\citenamefont {Ashcroft}(1968)}]{Ashcroft}%
  \BibitemOpen
  \bibfield  {author} {\bibinfo {author} {\bibfnamefont {N.~W.}\ \bibnamefont
  {Ashcroft}},\ }\href {\doibase 10.1103/PhysRevLett.21.1748} {\bibfield
  {journal} {\bibinfo  {journal} {Phys. Rev. Lett.}\ }\textbf {\bibinfo
  {volume} {21}},\ \bibinfo {pages} {1748} (\bibinfo {year}
  {1968})}\BibitemShut {NoStop}%
\bibitem [{\citenamefont {Ginzburg}(1969)}]{Ginzburg}%
  \BibitemOpen
  \bibfield  {author} {\bibinfo {author} {\bibfnamefont {V.~L.}\ \bibnamefont
  {Ginzburg}},\ }\href {\doibase 10.1070/PU1969v012n02ABEH003935} {\bibfield
  {journal} {\bibinfo  {journal} {Phys. Usp.}\ }\textbf {\bibinfo {volume}
  {12}},\ \bibinfo {pages} {241} (\bibinfo {year} {1969})}\BibitemShut
  {NoStop}%
\bibitem [{\citenamefont {Maksimov}\ and\ \citenamefont
  {Savrasov}(2001)}]{Maksimov}%
  \BibitemOpen
  \bibfield  {author} {\bibinfo {author} {\bibfnamefont {E.}~\bibnamefont
  {Maksimov}}\ and\ \bibinfo {author} {\bibfnamefont {D.}~\bibnamefont
  {Savrasov}},\ }\href {\doibase https://doi.org/10.1016/S0038-1098(01)00301-5}
  {\bibfield  {journal} {\bibinfo  {journal} {Solid State Commun.}\ }\textbf
  {\bibinfo {volume} {119}},\ \bibinfo {pages} {569 } (\bibinfo {year}
  {2001})}\BibitemShut {NoStop}%
\bibitem [{\citenamefont {Li}\ \emph {et~al.}(2014)\citenamefont {Li},
  \citenamefont {Hao}, \citenamefont {Liu}, \citenamefont {Li},\ and\
  \citenamefont {Ma}}]{Ma1}%
  \BibitemOpen
  \bibfield  {author} {\bibinfo {author} {\bibfnamefont {Y.}~\bibnamefont
  {Li}}, \bibinfo {author} {\bibfnamefont {J.}~\bibnamefont {Hao}}, \bibinfo
  {author} {\bibfnamefont {H.}~\bibnamefont {Liu}}, \bibinfo {author}
  {\bibfnamefont {Y.}~\bibnamefont {Li}}, \ and\ \bibinfo {author}
  {\bibfnamefont {Y.}~\bibnamefont {Ma}},\ }\href {\doibase 10.1063/1.4874158}
  {\bibfield  {journal} {\bibinfo  {journal} {J. Chem. Phys.}\ }\textbf
  {\bibinfo {volume} {140}},\ \bibinfo {pages} {174712} (\bibinfo {year}
  {2014})}\BibitemShut {NoStop}%
\bibitem [{\citenamefont {Duan}\ \emph {et~al.}(2014)\citenamefont {Duan},
  \citenamefont {Liu}, \citenamefont {Tian}, \citenamefont {Li}, \citenamefont
  {Huang}, \citenamefont {Zhao}, \citenamefont {Yu}, \citenamefont {Liu},
  \citenamefont {Tian},\ and\ \citenamefont {Cui}}]{Duan}%
  \BibitemOpen
  \bibfield  {author} {\bibinfo {author} {\bibfnamefont {D.}~\bibnamefont
  {Duan}}, \bibinfo {author} {\bibfnamefont {Y.}~\bibnamefont {Liu}}, \bibinfo
  {author} {\bibfnamefont {F.}~\bibnamefont {Tian}}, \bibinfo {author}
  {\bibfnamefont {D.}~\bibnamefont {Li}}, \bibinfo {author} {\bibfnamefont
  {X.}~\bibnamefont {Huang}}, \bibinfo {author} {\bibfnamefont
  {Z.}~\bibnamefont {Zhao}}, \bibinfo {author} {\bibfnamefont {H.}~\bibnamefont
  {Yu}}, \bibinfo {author} {\bibfnamefont {B.}~\bibnamefont {Liu}}, \bibinfo
  {author} {\bibfnamefont {W.}~\bibnamefont {Tian}}, \ and\ \bibinfo {author}
  {\bibfnamefont {T.}~\bibnamefont {Cui}},\ }\href
  {https://doi.org/10.1038/srep06968} {\bibfield  {journal} {\bibinfo
  {journal} {Sci. Rep.}\ }\textbf {\bibinfo {volume} {4}},\ \bibinfo {pages}
  {6968} (\bibinfo {year} {2014})}\BibitemShut {NoStop}%
\bibitem [{\citenamefont {Errea}\ \emph {et~al.}(2015)\citenamefont {Errea},
  \citenamefont {Calandra}, \citenamefont {Pickard}, \citenamefont {Nelson},
  \citenamefont {Needs}, \citenamefont {Li}, \citenamefont {Liu}, \citenamefont
  {Zhang}, \citenamefont {Ma},\ and\ \citenamefont {Mauri}}]{Errea_2015}%
  \BibitemOpen
  \bibfield  {author} {\bibinfo {author} {\bibfnamefont {I.}~\bibnamefont
  {Errea}}, \bibinfo {author} {\bibfnamefont {M.}~\bibnamefont {Calandra}},
  \bibinfo {author} {\bibfnamefont {C.~J.}\ \bibnamefont {Pickard}}, \bibinfo
  {author} {\bibfnamefont {J.}~\bibnamefont {Nelson}}, \bibinfo {author}
  {\bibfnamefont {R.~J.}\ \bibnamefont {Needs}}, \bibinfo {author}
  {\bibfnamefont {Y.}~\bibnamefont {Li}}, \bibinfo {author} {\bibfnamefont
  {H.}~\bibnamefont {Liu}}, \bibinfo {author} {\bibfnamefont {Y.}~\bibnamefont
  {Zhang}}, \bibinfo {author} {\bibfnamefont {Y.}~\bibnamefont {Ma}}, \ and\
  \bibinfo {author} {\bibfnamefont {F.}~\bibnamefont {Mauri}},\ }\href
  {\doibase 10.1103/PhysRevLett.114.157004} {\bibfield  {journal} {\bibinfo
  {journal} {Phys. Rev. Lett.}\ }\textbf {\bibinfo {volume} {114}},\ \bibinfo
  {pages} {157004} (\bibinfo {year} {2015})}\BibitemShut {NoStop}%
\bibitem [{\citenamefont {Einaga}\ \emph {et~al.}(2016)\citenamefont {Einaga},
  \citenamefont {Sakata}, \citenamefont {Ishikawa}, \citenamefont {Shimizu},
  \citenamefont {Eremets}, \citenamefont {Drozdov}, \citenamefont {Troyan},
  \citenamefont {Hirao},\ and\ \citenamefont {Ohishi}}]{Einaga}%
  \BibitemOpen
  \bibfield  {author} {\bibinfo {author} {\bibfnamefont {M.}~\bibnamefont
  {Einaga}}, \bibinfo {author} {\bibfnamefont {M.}~\bibnamefont {Sakata}},
  \bibinfo {author} {\bibfnamefont {T.}~\bibnamefont {Ishikawa}}, \bibinfo
  {author} {\bibfnamefont {K.}~\bibnamefont {Shimizu}}, \bibinfo {author}
  {\bibfnamefont {M.~I.}\ \bibnamefont {Eremets}}, \bibinfo {author}
  {\bibfnamefont {A.~P.}\ \bibnamefont {Drozdov}}, \bibinfo {author}
  {\bibfnamefont {I.~A.}\ \bibnamefont {Troyan}}, \bibinfo {author}
  {\bibfnamefont {N.}~\bibnamefont {Hirao}}, \ and\ \bibinfo {author}
  {\bibfnamefont {Y.}~\bibnamefont {Ohishi}},\ }\href
  {https://doi.org/10.1038/nphys3760} {\bibfield  {journal} {\bibinfo
  {journal} {Nat. Phys.}\ }\textbf {\bibinfo {volume} {12}},\ \bibinfo {pages}
  {835} (\bibinfo {year} {2016})}\BibitemShut {NoStop}%
\bibitem [{\citenamefont {Gor’kov}\ and\ \citenamefont
  {Kresin}(2016)}]{Kresin}%
  \BibitemOpen
  \bibfield  {author} {\bibinfo {author} {\bibfnamefont {L.~P.}\ \bibnamefont
  {Gor’kov}}\ and\ \bibinfo {author} {\bibfnamefont {V.~Z.}\ \bibnamefont
  {Kresin}},\ }\href {https://doi.org/10.1038/srep25608} {\bibfield  {journal}
  {\bibinfo  {journal} {Sci. Rep.}\ }\textbf {\bibinfo {volume} {6}},\ \bibinfo
  {pages} {25608} (\bibinfo {year} {2016})}\BibitemShut {NoStop}%
\bibitem [{\citenamefont {{Flores-Livas, Jos\'eA.}}\ \emph
  {et~al.}(2016)\citenamefont {{Flores-Livas, Jos\'eA.}}, \citenamefont
  {{Sanna, Antonio}},\ and\ \citenamefont {{Gross, E. K.U.}}}]{phonon1}%
  \BibitemOpen
  \bibfield  {author} {\bibinfo {author} {\bibnamefont {{Flores-Livas,
  Jos\'eA.}}}, \bibinfo {author} {\bibnamefont {{Sanna, Antonio}}}, \ and\
  \bibinfo {author} {\bibnamefont {{Gross, E. K.U.}}},\ }\href {\doibase
  10.1140/epjb/e2016-70020-0} {\bibfield  {journal} {\bibinfo  {journal} {Eur.
  Phys. J. B}\ }\textbf {\bibinfo {volume} {89}},\ \bibinfo {pages} {63}
  (\bibinfo {year} {2016})}\BibitemShut {NoStop}%
\bibitem [{\citenamefont {Akashi}\ \emph {et~al.}(2015)\citenamefont {Akashi},
  \citenamefont {Kawamura}, \citenamefont {Tsuneyuki}, \citenamefont {Nomura},\
  and\ \citenamefont {Arita}}]{phonon2}%
  \BibitemOpen
  \bibfield  {author} {\bibinfo {author} {\bibfnamefont {R.}~\bibnamefont
  {Akashi}}, \bibinfo {author} {\bibfnamefont {M.}~\bibnamefont {Kawamura}},
  \bibinfo {author} {\bibfnamefont {S.}~\bibnamefont {Tsuneyuki}}, \bibinfo
  {author} {\bibfnamefont {Y.}~\bibnamefont {Nomura}}, \ and\ \bibinfo {author}
  {\bibfnamefont {R.}~\bibnamefont {Arita}},\ }\href {\doibase
  10.1103/PhysRevB.91.224513} {\bibfield  {journal} {\bibinfo  {journal} {Phys.
  Rev. B}\ }\textbf {\bibinfo {volume} {91}},\ \bibinfo {pages} {224513}
  (\bibinfo {year} {2015})}\BibitemShut {NoStop}%
\bibitem [{\citenamefont {Gor'kov}\ and\ \citenamefont
  {Kresin}(2018)}]{Kresin2}%
  \BibitemOpen
  \bibfield  {author} {\bibinfo {author} {\bibfnamefont {L.~P.}\ \bibnamefont
  {Gor'kov}}\ and\ \bibinfo {author} {\bibfnamefont {V.~Z.}\ \bibnamefont
  {Kresin}},\ }\href {\doibase 10.1103/RevModPhys.90.011001} {\bibfield
  {journal} {\bibinfo  {journal} {Rev. Mod. Phys.}\ }\textbf {\bibinfo {volume}
  {90}},\ \bibinfo {pages} {011001} (\bibinfo {year} {2018})}\BibitemShut
  {NoStop}%
\bibitem [{\citenamefont {Bernstein}\ \emph {et~al.}(2015)\citenamefont
  {Bernstein}, \citenamefont {Hellberg}, \citenamefont {Johannes},
  \citenamefont {Mazin},\ and\ \citenamefont {Mehl}}]{Igor}%
  \BibitemOpen
  \bibfield  {author} {\bibinfo {author} {\bibfnamefont {N.}~\bibnamefont
  {Bernstein}}, \bibinfo {author} {\bibfnamefont {C.~S.}\ \bibnamefont
  {Hellberg}}, \bibinfo {author} {\bibfnamefont {M.~D.}\ \bibnamefont
  {Johannes}}, \bibinfo {author} {\bibfnamefont {I.~I.}\ \bibnamefont {Mazin}},
  \ and\ \bibinfo {author} {\bibfnamefont {M.~J.}\ \bibnamefont {Mehl}},\
  }\href {\doibase 10.1103/PhysRevB.91.060511} {\bibfield  {journal} {\bibinfo
  {journal} {Phys. Rev. B}\ }\textbf {\bibinfo {volume} {91}},\ \bibinfo
  {pages} {060511} (\bibinfo {year} {2015})}\BibitemShut {NoStop}%
\bibitem [{\citenamefont {Quan}\ and\ \citenamefont {Pickett}(2016)}]{Pickett}%
  \BibitemOpen
  \bibfield  {author} {\bibinfo {author} {\bibfnamefont {Y.}~\bibnamefont
  {Quan}}\ and\ \bibinfo {author} {\bibfnamefont {W.~E.}\ \bibnamefont
  {Pickett}},\ }\href {\doibase 10.1103/PhysRevB.93.104526} {\bibfield
  {journal} {\bibinfo  {journal} {Phys. Rev. B}\ }\textbf {\bibinfo {volume}
  {93}},\ \bibinfo {pages} {104526} (\bibinfo {year} {2016})}\BibitemShut
  {NoStop}%
\bibitem [{\citenamefont {Bianconi}\ and\ \citenamefont
  {Jarlborg}(2015)}]{Bianconi}%
  \BibitemOpen
  \bibfield  {author} {\bibinfo {author} {\bibfnamefont {A.}~\bibnamefont
  {Bianconi}}\ and\ \bibinfo {author} {\bibfnamefont {T.}~\bibnamefont
  {Jarlborg}},\ }\href {http://stacks.iop.org/0295-5075/112/i=3/a=37001}
  {\bibfield  {journal} {\bibinfo  {journal} {EPL (Europhysics Letters)}\
  }\textbf {\bibinfo {volume} {112}},\ \bibinfo {pages} {37001} (\bibinfo
  {year} {2015})}\BibitemShut {NoStop}%
\bibitem [{\citenamefont {Heil}\ and\ \citenamefont {Boeri}(2015)}]{Heil}%
  \BibitemOpen
  \bibfield  {author} {\bibinfo {author} {\bibfnamefont {C.}~\bibnamefont
  {Heil}}\ and\ \bibinfo {author} {\bibfnamefont {L.}~\bibnamefont {Boeri}},\
  }\href {\doibase 10.1103/PhysRevB.92.060508} {\bibfield  {journal} {\bibinfo
  {journal} {Phys. Rev. B}\ }\textbf {\bibinfo {volume} {92}},\ \bibinfo
  {pages} {060508} (\bibinfo {year} {2015})}\BibitemShut {NoStop}%
\bibitem [{\citenamefont {Flores-Livas}\ \emph {et~al.}(2016)\citenamefont
  {Flores-Livas}, \citenamefont {Amsler}, \citenamefont {Heil}, \citenamefont
  {Sanna}, \citenamefont {Boeri}, \citenamefont {Profeta}, \citenamefont
  {Wolverton}, \citenamefont {Goedecker},\ and\ \citenamefont
  {Gross}}]{Flores}%
  \BibitemOpen
  \bibfield  {author} {\bibinfo {author} {\bibfnamefont {J.~A.}\ \bibnamefont
  {Flores-Livas}}, \bibinfo {author} {\bibfnamefont {M.}~\bibnamefont
  {Amsler}}, \bibinfo {author} {\bibfnamefont {C.}~\bibnamefont {Heil}},
  \bibinfo {author} {\bibfnamefont {A.}~\bibnamefont {Sanna}}, \bibinfo
  {author} {\bibfnamefont {L.}~\bibnamefont {Boeri}}, \bibinfo {author}
  {\bibfnamefont {G.}~\bibnamefont {Profeta}}, \bibinfo {author} {\bibfnamefont
  {C.}~\bibnamefont {Wolverton}}, \bibinfo {author} {\bibfnamefont
  {S.}~\bibnamefont {Goedecker}}, \ and\ \bibinfo {author} {\bibfnamefont
  {E.~K.~U.}\ \bibnamefont {Gross}},\ }\href {\doibase
  10.1103/PhysRevB.93.020508} {\bibfield  {journal} {\bibinfo  {journal} {Phys.
  Rev. B}\ }\textbf {\bibinfo {volume} {93}},\ \bibinfo {pages} {020508}
  (\bibinfo {year} {2016})}\BibitemShut {NoStop}%
\bibitem [{\citenamefont {Abrikosov}(2004)}]{Abrikosov}%
  \BibitemOpen
  \bibfield  {author} {\bibinfo {author} {\bibfnamefont {A.~A.}\ \bibnamefont
  {Abrikosov}},\ }\href {\doibase 10.1103/RevModPhys.76.975} {\bibfield
  {journal} {\bibinfo  {journal} {Rev. Mod. Phys.}\ }\textbf {\bibinfo {volume}
  {76}},\ \bibinfo {pages} {975} (\bibinfo {year} {2004})}\BibitemShut
  {NoStop}%
\bibitem [{\citenamefont {Clogston}(1962)}]{Clogston}%
  \BibitemOpen
  \bibfield  {author} {\bibinfo {author} {\bibfnamefont {A.~M.}\ \bibnamefont
  {Clogston}},\ }\href {\doibase 10.1103/PhysRevLett.9.266} {\bibfield
  {journal} {\bibinfo  {journal} {Phys. Rev. Lett.}\ }\textbf {\bibinfo
  {volume} {9}},\ \bibinfo {pages} {266} (\bibinfo {year} {1962})}\BibitemShut
  {NoStop}%
\bibitem [{\citenamefont {Chandrasekhar}(1962)}]{Chandrasekhar}%
  \BibitemOpen
  \bibfield  {author} {\bibinfo {author} {\bibfnamefont {B.~S.}\ \bibnamefont
  {Chandrasekhar}},\ }\href {\doibase 10.1063/1.1777362} {\bibfield  {journal}
  {\bibinfo  {journal} {Appl. Phys. Lett.}\ }\textbf {\bibinfo {volume} {1}},\
  \bibinfo {pages} {7} (\bibinfo {year} {1962})}\BibitemShut {NoStop}%
\bibitem [{\citenamefont {Maki}(1966)}]{Maki}%
  \BibitemOpen
  \bibfield  {author} {\bibinfo {author} {\bibfnamefont {K.}~\bibnamefont
  {Maki}},\ }\href {\doibase 10.1103/PhysRev.148.362} {\bibfield  {journal}
  {\bibinfo  {journal} {Phys. Rev.}\ }\textbf {\bibinfo {volume} {148}},\
  \bibinfo {pages} {362} (\bibinfo {year} {1966})}\BibitemShut {NoStop}%
\bibitem [{\citenamefont {Werthamer}\ \emph {et~al.}(1966)\citenamefont
  {Werthamer}, \citenamefont {Helfand},\ and\ \citenamefont {Hohenberg}}]{WHH}%
  \BibitemOpen
  \bibfield  {author} {\bibinfo {author} {\bibfnamefont {N.~R.}\ \bibnamefont
  {Werthamer}}, \bibinfo {author} {\bibfnamefont {E.}~\bibnamefont {Helfand}},
  \ and\ \bibinfo {author} {\bibfnamefont {P.~C.}\ \bibnamefont {Hohenberg}},\
  }\href {\doibase 10.1103/PhysRev.147.295} {\bibfield  {journal} {\bibinfo
  {journal} {Phys. Rev.}\ }\textbf {\bibinfo {volume} {147}},\ \bibinfo {pages}
  {295} (\bibinfo {year} {1966})}\BibitemShut {NoStop}%
\bibitem [{\citenamefont {Braithwaite}\ \emph {et~al.}(2016)\citenamefont
  {Braithwaite}, \citenamefont {Knafo}, \citenamefont {Settai}, \citenamefont
  {Aoki}, \citenamefont {Kurahashi},\ and\ \citenamefont {Flouquet}}]{Jacques}%
  \BibitemOpen
  \bibfield  {author} {\bibinfo {author} {\bibfnamefont {D.}~\bibnamefont
  {Braithwaite}}, \bibinfo {author} {\bibfnamefont {W.}~\bibnamefont {Knafo}},
  \bibinfo {author} {\bibfnamefont {R.}~\bibnamefont {Settai}}, \bibinfo
  {author} {\bibfnamefont {D.}~\bibnamefont {Aoki}}, \bibinfo {author}
  {\bibfnamefont {S.}~\bibnamefont {Kurahashi}}, \ and\ \bibinfo {author}
  {\bibfnamefont {J.}~\bibnamefont {Flouquet}},\ }\href {\doibase
  10.1063/1.4941714} {\bibfield  {journal} {\bibinfo  {journal} {Rev. Sci.
  Instrum.}\ }\textbf {\bibinfo {volume} {87}},\ \bibinfo {pages} {023907}
  (\bibinfo {year} {2016})}\BibitemShut {NoStop}%
\bibitem [{\citenamefont {Eremets}(2003)}]{ME-Raman}%
  \BibitemOpen
  \bibfield  {author} {\bibinfo {author} {\bibfnamefont {M.~I.}\ \bibnamefont
  {Eremets}},\ }\href {\doibase 10.1002/jrs.1044} {\bibfield  {journal}
  {\bibinfo  {journal} {J. Raman Spectrosc.}\ }\textbf {\bibinfo {volume}
  {34}},\ \bibinfo {pages} {515} (\bibinfo {year} {2003})}\BibitemShut
  {NoStop}%
\bibitem [{\citenamefont {Eremets}\ and\ \citenamefont
  {Troyan}(2011)}]{ME-dense}%
  \BibitemOpen
  \bibfield  {author} {\bibinfo {author} {\bibfnamefont {M.~I.}\ \bibnamefont
  {Eremets}}\ and\ \bibinfo {author} {\bibfnamefont {I.~A.}\ \bibnamefont
  {Troyan}},\ }\href {https://doi.org/10.1038/nmat3175} {\bibfield  {journal}
  {\bibinfo  {journal} {Nat. Mater.}\ }\textbf {\bibinfo {volume} {10}},\
  \bibinfo {pages} {927} (\bibinfo {year} {2011})}\BibitemShut {NoStop}%
\bibitem [{\citenamefont {Meier}\ \emph {et~al.}(2016)\citenamefont {Meier},
  \citenamefont {Kong}, \citenamefont {Kaluarachchi}, \citenamefont {Taufour},
  \citenamefont {Jo}, \citenamefont {Drachuck}, \citenamefont {B\"ohmer},
  \citenamefont {Saunders}, \citenamefont {Sapkota}, \citenamefont {Kreyssig},
  \citenamefont {Tanatar}, \citenamefont {Prozorov}, \citenamefont {Goldman},
  \citenamefont {Balakirev}, \citenamefont {Gurevich}, \citenamefont {Bud'ko},\
  and\ \citenamefont {Canfield}}]{Budko}%
  \BibitemOpen
  \bibfield  {author} {\bibinfo {author} {\bibfnamefont {W.~R.}\ \bibnamefont
  {Meier}}, \bibinfo {author} {\bibfnamefont {T.}~\bibnamefont {Kong}},
  \bibinfo {author} {\bibfnamefont {U.~S.}\ \bibnamefont {Kaluarachchi}},
  \bibinfo {author} {\bibfnamefont {V.}~\bibnamefont {Taufour}}, \bibinfo
  {author} {\bibfnamefont {N.~H.}\ \bibnamefont {Jo}}, \bibinfo {author}
  {\bibfnamefont {G.}~\bibnamefont {Drachuck}}, \bibinfo {author}
  {\bibfnamefont {A.~E.}\ \bibnamefont {B\"ohmer}}, \bibinfo {author}
  {\bibfnamefont {S.~M.}\ \bibnamefont {Saunders}}, \bibinfo {author}
  {\bibfnamefont {A.}~\bibnamefont {Sapkota}}, \bibinfo {author} {\bibfnamefont
  {A.}~\bibnamefont {Kreyssig}}, \bibinfo {author} {\bibfnamefont {M.~A.}\
  \bibnamefont {Tanatar}}, \bibinfo {author} {\bibfnamefont {R.}~\bibnamefont
  {Prozorov}}, \bibinfo {author} {\bibfnamefont {A.~I.}\ \bibnamefont
  {Goldman}}, \bibinfo {author} {\bibfnamefont {F.~F.}\ \bibnamefont
  {Balakirev}}, \bibinfo {author} {\bibfnamefont {A.}~\bibnamefont {Gurevich}},
  \bibinfo {author} {\bibfnamefont {S.~L.}\ \bibnamefont {Bud'ko}}, \ and\
  \bibinfo {author} {\bibfnamefont {P.~C.}\ \bibnamefont {Canfield}},\ }\href
  {\doibase 10.1103/PhysRevB.94.064501} {\bibfield  {journal} {\bibinfo
  {journal} {Phys. Rev. B}\ }\textbf {\bibinfo {volume} {94}},\ \bibinfo
  {pages} {064501} (\bibinfo {year} {2016})}\BibitemShut {NoStop}%
\bibitem [{\citenamefont {Mielke}\ \emph {et~al.}(2001)\citenamefont {Mielke},
  \citenamefont {Singleton}, \citenamefont {Nam}, \citenamefont {Harrison},
  \citenamefont {Agosta}, \citenamefont {Fravel},\ and\ \citenamefont
  {Montgomery}}]{Mielke_2001}%
  \BibitemOpen
  \bibfield  {author} {\bibinfo {author} {\bibfnamefont {C.}~\bibnamefont
  {Mielke}}, \bibinfo {author} {\bibfnamefont {J.}~\bibnamefont {Singleton}},
  \bibinfo {author} {\bibfnamefont {M.-S.}\ \bibnamefont {Nam}}, \bibinfo
  {author} {\bibfnamefont {N.}~\bibnamefont {Harrison}}, \bibinfo {author}
  {\bibfnamefont {C.~C.}\ \bibnamefont {Agosta}}, \bibinfo {author}
  {\bibfnamefont {B.}~\bibnamefont {Fravel}}, \ and\ \bibinfo {author}
  {\bibfnamefont {L.~K.}\ \bibnamefont {Montgomery}},\ }\href {\doibase
  10.1088/0953-8984/13/36/308} {\bibfield  {journal} {\bibinfo  {journal} {J.
  Phys. Condens. Matter}\ }\textbf {\bibinfo {volume} {13}},\ \bibinfo {pages}
  {8325} (\bibinfo {year} {2001})}\BibitemShut {NoStop}%
\bibitem [{\citenamefont {Altarawneh}\ \emph {et~al.}(2008)\citenamefont
  {Altarawneh}, \citenamefont {Collar}, \citenamefont {Mielke}, \citenamefont
  {Ni}, \citenamefont {Bud'ko},\ and\ \citenamefont
  {Canfield}}]{Altarawneh_2008}%
  \BibitemOpen
  \bibfield  {author} {\bibinfo {author} {\bibfnamefont {M.~M.}\ \bibnamefont
  {Altarawneh}}, \bibinfo {author} {\bibfnamefont {K.}~\bibnamefont {Collar}},
  \bibinfo {author} {\bibfnamefont {C.~H.}\ \bibnamefont {Mielke}}, \bibinfo
  {author} {\bibfnamefont {N.}~\bibnamefont {Ni}}, \bibinfo {author}
  {\bibfnamefont {S.~L.}\ \bibnamefont {Bud'ko}}, \ and\ \bibinfo {author}
  {\bibfnamefont {P.~C.}\ \bibnamefont {Canfield}},\ }\href {\doibase
  10.1103/PhysRevB.78.220505} {\bibfield  {journal} {\bibinfo  {journal} {Phys.
  Rev. B}\ }\textbf {\bibinfo {volume} {78}},\ \bibinfo {pages} {220505}
  (\bibinfo {year} {2008})}\BibitemShut {NoStop}%
\bibitem [{\citenamefont {Grissonnanche}\ \emph {et~al.}(2014)\citenamefont
  {Grissonnanche}, \citenamefont {Cyr-Choini{\`e}re}, \citenamefont
  {Lalibert{\'e}}, \citenamefont {Ren{\'e}~de Cotret}, \citenamefont
  {Juneau-Fecteau}, \citenamefont {Dufour-Beaus{\'e}jour}, \citenamefont
  {Delage}, \citenamefont {LeBoeuf}, \citenamefont {Chang}, \citenamefont
  {Ramshaw}, \citenamefont {Bonn}, \citenamefont {Hardy}, \citenamefont
  {Liang}, \citenamefont {Adachi}, \citenamefont {Hussey}, \citenamefont
  {Vignolle}, \citenamefont {Proust}, \citenamefont {Sutherland}, \citenamefont
  {Kr{\"a}mer}, \citenamefont {Park}, \citenamefont {Graf}, \citenamefont
  {Doiron-Leyraud},\ and\ \citenamefont {Taillefer}}]{Grissonnanche2014}%
  \BibitemOpen
  \bibfield  {author} {\bibinfo {author} {\bibfnamefont {G.}~\bibnamefont
  {Grissonnanche}}, \bibinfo {author} {\bibfnamefont {O.}~\bibnamefont
  {Cyr-Choini{\`e}re}}, \bibinfo {author} {\bibfnamefont {F.}~\bibnamefont
  {Lalibert{\'e}}}, \bibinfo {author} {\bibfnamefont {S.}~\bibnamefont
  {Ren{\'e}~de Cotret}}, \bibinfo {author} {\bibfnamefont {A.}~\bibnamefont
  {Juneau-Fecteau}}, \bibinfo {author} {\bibfnamefont {S.}~\bibnamefont
  {Dufour-Beaus{\'e}jour}}, \bibinfo {author} {\bibfnamefont {M.-{\`E}.}\
  \bibnamefont {Delage}}, \bibinfo {author} {\bibfnamefont {D.}~\bibnamefont
  {LeBoeuf}}, \bibinfo {author} {\bibfnamefont {J.}~\bibnamefont {Chang}},
  \bibinfo {author} {\bibfnamefont {B.~J.}\ \bibnamefont {Ramshaw}}, \bibinfo
  {author} {\bibfnamefont {D.~A.}\ \bibnamefont {Bonn}}, \bibinfo {author}
  {\bibfnamefont {W.~N.}\ \bibnamefont {Hardy}}, \bibinfo {author}
  {\bibfnamefont {R.}~\bibnamefont {Liang}}, \bibinfo {author} {\bibfnamefont
  {S.}~\bibnamefont {Adachi}}, \bibinfo {author} {\bibfnamefont {N.~E.}\
  \bibnamefont {Hussey}}, \bibinfo {author} {\bibfnamefont {B.}~\bibnamefont
  {Vignolle}}, \bibinfo {author} {\bibfnamefont {C.}~\bibnamefont {Proust}},
  \bibinfo {author} {\bibfnamefont {M.}~\bibnamefont {Sutherland}}, \bibinfo
  {author} {\bibfnamefont {S.}~\bibnamefont {Kr{\"a}mer}}, \bibinfo {author}
  {\bibfnamefont {J.-H.}\ \bibnamefont {Park}}, \bibinfo {author}
  {\bibfnamefont {D.}~\bibnamefont {Graf}}, \bibinfo {author} {\bibfnamefont
  {N.}~\bibnamefont {Doiron-Leyraud}}, \ and\ \bibinfo {author} {\bibfnamefont
  {L.}~\bibnamefont {Taillefer}},\ }\href {https://doi.org/10.1038/ncomms4280}
  {\bibfield  {journal} {\bibinfo  {journal} {Nat. Commun.}\ }\textbf {\bibinfo
  {volume} {5}},\ \bibinfo {pages} {3280 EP } (\bibinfo {year}
  {2014})}\BibitemShut {NoStop}%
\bibitem [{\citenamefont {Bulaevskii}\ \emph {et~al.}(1988)\citenamefont
  {Bulaevskii}, \citenamefont {Dolgov},\ and\ \citenamefont
  {Ptitsyn}}]{Boulaevskii_1988}%
  \BibitemOpen
  \bibfield  {author} {\bibinfo {author} {\bibfnamefont {L.~N.}\ \bibnamefont
  {Bulaevskii}}, \bibinfo {author} {\bibfnamefont {O.~V.}\ \bibnamefont
  {Dolgov}}, \ and\ \bibinfo {author} {\bibfnamefont {M.~O.}\ \bibnamefont
  {Ptitsyn}},\ }\href {\doibase 10.1103/PhysRevB.38.11290} {\bibfield
  {journal} {\bibinfo  {journal} {Phys. Rev. B}\ }\textbf {\bibinfo {volume}
  {38}},\ \bibinfo {pages} {11290} (\bibinfo {year} {1988})}\BibitemShut
  {NoStop}%
\bibitem [{\citenamefont {Thomas}\ \emph {et~al.}(1996)\citenamefont {Thomas},
  \citenamefont {Wand}, \citenamefont {L{\"u}hmann}, \citenamefont {Gegenwart},
  \citenamefont {Stewart}, \citenamefont {Steglich}, \citenamefont {Brison},
  \citenamefont {Buzdin}, \citenamefont {Gl{\'e}mot},\ and\ \citenamefont
  {Flouquet}}]{Thomas1996}%
  \BibitemOpen
  \bibfield  {author} {\bibinfo {author} {\bibfnamefont {F.}~\bibnamefont
  {Thomas}}, \bibinfo {author} {\bibfnamefont {B.}~\bibnamefont {Wand}},
  \bibinfo {author} {\bibfnamefont {T.}~\bibnamefont {L{\"u}hmann}}, \bibinfo
  {author} {\bibfnamefont {P.}~\bibnamefont {Gegenwart}}, \bibinfo {author}
  {\bibfnamefont {G.~R.}\ \bibnamefont {Stewart}}, \bibinfo {author}
  {\bibfnamefont {F.}~\bibnamefont {Steglich}}, \bibinfo {author}
  {\bibfnamefont {J.~P.}\ \bibnamefont {Brison}}, \bibinfo {author}
  {\bibfnamefont {A.}~\bibnamefont {Buzdin}}, \bibinfo {author} {\bibfnamefont
  {L.}~\bibnamefont {Gl{\'e}mot}}, \ and\ \bibinfo {author} {\bibfnamefont
  {J.}~\bibnamefont {Flouquet}},\ }\href {\doibase 10.1007/BF00755113}
  {\bibfield  {journal} {\bibinfo  {journal} {J. Low Temp. Phys.}\ }\textbf
  {\bibinfo {volume} {102}},\ \bibinfo {pages} {117} (\bibinfo {year}
  {1996})}\BibitemShut {NoStop}%
\bibitem [{\citenamefont {Lee}\ \emph {et~al.}(2001)\citenamefont {Lee},
  \citenamefont {Mori}, \citenamefont {Masui}, \citenamefont {Eltsev},
  \citenamefont {Yamamoto},\ and\ \citenamefont {Tajima}}]{MgB2}%
  \BibitemOpen
  \bibfield  {author} {\bibinfo {author} {\bibfnamefont {S.}~\bibnamefont
  {Lee}}, \bibinfo {author} {\bibfnamefont {H.}~\bibnamefont {Mori}}, \bibinfo
  {author} {\bibfnamefont {T.}~\bibnamefont {Masui}}, \bibinfo {author}
  {\bibfnamefont {Y.}~\bibnamefont {Eltsev}}, \bibinfo {author} {\bibfnamefont
  {A.}~\bibnamefont {Yamamoto}}, \ and\ \bibinfo {author} {\bibfnamefont
  {S.}~\bibnamefont {Tajima}},\ }\href {\doibase 10.1143/JPSJ.70.2255}
  {\bibfield  {journal} {\bibinfo  {journal} {J. Physical. Soc. Japan.}\
  }\textbf {\bibinfo {volume} {70}},\ \bibinfo {pages} {2255} (\bibinfo {year}
  {2001})}\BibitemShut {NoStop}%
\bibitem [{\citenamefont {Gurevich}(2003)}]{AlexG}%
  \BibitemOpen
  \bibfield  {author} {\bibinfo {author} {\bibfnamefont {A.}~\bibnamefont
  {Gurevich}},\ }\href {\doibase 10.1103/PhysRevB.67.184515} {\bibfield
  {journal} {\bibinfo  {journal} {Phys. Rev. B}\ }\textbf {\bibinfo {volume}
  {67}},\ \bibinfo {pages} {184515} (\bibinfo {year} {2003})}\BibitemShut
  {NoStop}%
\bibitem [{\citenamefont {Zhang}\ \emph {et~al.}(2016)\citenamefont {Zhang},
  \citenamefont {Rhodes}, \citenamefont {Zeng}, \citenamefont {Johannes},\ and\
  \citenamefont {Balicas}}]{Luis-non-Ising}%
  \BibitemOpen
  \bibfield  {author} {\bibinfo {author} {\bibfnamefont {Q.~R.}\ \bibnamefont
  {Zhang}}, \bibinfo {author} {\bibfnamefont {D.}~\bibnamefont {Rhodes}},
  \bibinfo {author} {\bibfnamefont {B.}~\bibnamefont {Zeng}}, \bibinfo {author}
  {\bibfnamefont {M.~D.}\ \bibnamefont {Johannes}}, \ and\ \bibinfo {author}
  {\bibfnamefont {L.}~\bibnamefont {Balicas}},\ }\href {\doibase
  10.1103/PhysRevB.94.094511} {\bibfield  {journal} {\bibinfo  {journal} {Phys.
  Rev. B}\ }\textbf {\bibinfo {volume} {94}},\ \bibinfo {pages} {094511}
  (\bibinfo {year} {2016})}\BibitemShut {NoStop}%
\bibitem [{\citenamefont {Jaroszynski}\ \emph {et~al.}(2008)\citenamefont
  {Jaroszynski}, \citenamefont {Riggs}, \citenamefont {Hunte}, \citenamefont
  {Gurevich}, \citenamefont {Larbalestier}, \citenamefont {Boebinger},
  \citenamefont {Balakirev}, \citenamefont {Migliori}, \citenamefont {Ren},
  \citenamefont {Lu}, \citenamefont {Yang}, \citenamefont {Shen}, \citenamefont
  {Dong}, \citenamefont {Zhao}, \citenamefont {Jin}, \citenamefont {Sefat},
  \citenamefont {McGuire}, \citenamefont {Sales}, \citenamefont {Christen},\
  and\ \citenamefont {Mandrus}}]{FFB}%
  \BibitemOpen
  \bibfield  {author} {\bibinfo {author} {\bibfnamefont {J.}~\bibnamefont
  {Jaroszynski}}, \bibinfo {author} {\bibfnamefont {S.~C.}\ \bibnamefont
  {Riggs}}, \bibinfo {author} {\bibfnamefont {F.}~\bibnamefont {Hunte}},
  \bibinfo {author} {\bibfnamefont {A.}~\bibnamefont {Gurevich}}, \bibinfo
  {author} {\bibfnamefont {D.~C.}\ \bibnamefont {Larbalestier}}, \bibinfo
  {author} {\bibfnamefont {G.~S.}\ \bibnamefont {Boebinger}}, \bibinfo {author}
  {\bibfnamefont {F.~F.}\ \bibnamefont {Balakirev}}, \bibinfo {author}
  {\bibfnamefont {A.}~\bibnamefont {Migliori}}, \bibinfo {author}
  {\bibfnamefont {Z.~A.}\ \bibnamefont {Ren}}, \bibinfo {author} {\bibfnamefont
  {W.}~\bibnamefont {Lu}}, \bibinfo {author} {\bibfnamefont {J.}~\bibnamefont
  {Yang}}, \bibinfo {author} {\bibfnamefont {X.~L.}\ \bibnamefont {Shen}},
  \bibinfo {author} {\bibfnamefont {X.~L.}\ \bibnamefont {Dong}}, \bibinfo
  {author} {\bibfnamefont {Z.~X.}\ \bibnamefont {Zhao}}, \bibinfo {author}
  {\bibfnamefont {R.}~\bibnamefont {Jin}}, \bibinfo {author} {\bibfnamefont
  {A.~S.}\ \bibnamefont {Sefat}}, \bibinfo {author} {\bibfnamefont {M.~A.}\
  \bibnamefont {McGuire}}, \bibinfo {author} {\bibfnamefont {B.~C.}\
  \bibnamefont {Sales}}, \bibinfo {author} {\bibfnamefont {D.~K.}\ \bibnamefont
  {Christen}}, \ and\ \bibinfo {author} {\bibfnamefont {D.}~\bibnamefont
  {Mandrus}},\ }\href {\doibase 10.1103/PhysRevB.78.064511} {\bibfield
  {journal} {\bibinfo  {journal} {Phys. Rev. B}\ }\textbf {\bibinfo {volume}
  {78}},\ \bibinfo {pages} {064511} (\bibinfo {year} {2008})}\BibitemShut
  {NoStop}%
\bibitem [{\citenamefont {Yuan}\ \emph {et~al.}(2009)\citenamefont {Yuan},
  \citenamefont {Singleton}, \citenamefont {Balakirev}, \citenamefont {Baily},
  \citenamefont {Chen}, \citenamefont {Luo},\ and\ \citenamefont
  {Wang}}]{NLWang}%
  \BibitemOpen
  \bibfield  {author} {\bibinfo {author} {\bibfnamefont {H.~Q.}\ \bibnamefont
  {Yuan}}, \bibinfo {author} {\bibfnamefont {J.}~\bibnamefont {Singleton}},
  \bibinfo {author} {\bibfnamefont {F.~F.}\ \bibnamefont {Balakirev}}, \bibinfo
  {author} {\bibfnamefont {S.~A.}\ \bibnamefont {Baily}}, \bibinfo {author}
  {\bibfnamefont {G.~F.}\ \bibnamefont {Chen}}, \bibinfo {author}
  {\bibfnamefont {J.~L.}\ \bibnamefont {Luo}}, \ and\ \bibinfo {author}
  {\bibfnamefont {N.~L.}\ \bibnamefont {Wang}},\ }\href {\doibase
  10.1038/nature07676} {\bibfield  {journal} {\bibinfo  {journal} {Nature}\
  }\textbf {\bibinfo {volume} {459}},\ \bibinfo {pages} {565} (\bibinfo {year}
  {2009})}\BibitemShut {NoStop}%
\bibitem [{\citenamefont {Errea}\ \emph {et~al.}(2016)\citenamefont {Errea},
  \citenamefont {Calandra}, \citenamefont {Pickard}, \citenamefont {Nelson},
  \citenamefont {Needs}, \citenamefont {Li}, \citenamefont {Liu}, \citenamefont
  {Zhang}, \citenamefont {Ma},\ and\ \citenamefont {Mauri}}]{Errea2016}%
  \BibitemOpen
  \bibfield  {author} {\bibinfo {author} {\bibfnamefont {I.}~\bibnamefont
  {Errea}}, \bibinfo {author} {\bibfnamefont {M.}~\bibnamefont {Calandra}},
  \bibinfo {author} {\bibfnamefont {C.~J.}\ \bibnamefont {Pickard}}, \bibinfo
  {author} {\bibfnamefont {J.~R.}\ \bibnamefont {Nelson}}, \bibinfo {author}
  {\bibfnamefont {R.~J.}\ \bibnamefont {Needs}}, \bibinfo {author}
  {\bibfnamefont {Y.}~\bibnamefont {Li}}, \bibinfo {author} {\bibfnamefont
  {H.}~\bibnamefont {Liu}}, \bibinfo {author} {\bibfnamefont {Y.}~\bibnamefont
  {Zhang}}, \bibinfo {author} {\bibfnamefont {Y.}~\bibnamefont {Ma}}, \ and\
  \bibinfo {author} {\bibfnamefont {F.}~\bibnamefont {Mauri}},\ }\href
  {https://doi.org/10.1038/nature17175} {\bibfield  {journal} {\bibinfo
  {journal} {Nature}\ }\textbf {\bibinfo {volume} {532}},\ \bibinfo {pages}
  {81} (\bibinfo {year} {2016})}\BibitemShut {NoStop}%
\bibitem [{\citenamefont {Brown}\ \emph {et~al.}(2018)\citenamefont {Brown},
  \citenamefont {Semeniuk}, \citenamefont {Wang}, \citenamefont {Monserrat},
  \citenamefont {Pickard},\ and\ \citenamefont {Grosche}}]{Brown_2018}%
  \BibitemOpen
  \bibfield  {author} {\bibinfo {author} {\bibfnamefont {P.}~\bibnamefont
  {Brown}}, \bibinfo {author} {\bibfnamefont {K.}~\bibnamefont {Semeniuk}},
  \bibinfo {author} {\bibfnamefont {D.}~\bibnamefont {Wang}}, \bibinfo {author}
  {\bibfnamefont {B.}~\bibnamefont {Monserrat}}, \bibinfo {author}
  {\bibfnamefont {C.~J.}\ \bibnamefont {Pickard}}, \ and\ \bibinfo {author}
  {\bibfnamefont {F.~M.}\ \bibnamefont {Grosche}},\ }\href {\doibase
  10.1126/sciadv.aao4793} {\bibfield  {journal} {\bibinfo  {journal} {Sci.
  Adv.}\ }\textbf {\bibinfo {volume} {4}} (\bibinfo {year} {2018}),\
  10.1126/sciadv.aao4793}\BibitemShut {NoStop}%
\end{thebibliography}%

\end{document}